\documentstyle[epsf,prb,twocolumn,floats,aps]{revtex}

\begin{document}

\twocolumn[\hsize\textwidth\columnwidth\hsize\csname
@twocolumnfalse\endcsname
 \draft
\title{Superfluid transition temperature from the Lindemann-like criterion}
\author{S.M. Apenko$^{\ast}$}
\address{I.E. Tamm Theory Division, P.N. Lebedev Physical Institute, Moscow, 117924,
Russia}
\date{\today }
\maketitle

\begin{abstract}
We further analyze the recently proposed criterion, according to
which a superfluid transition occurs whenever the r.m.s.
displacement of a particle from its initial position in imaginary
time reaches a certain fraction of the interparticle distance. The
critical temperature $T_\lambda$ is expressed in terms of the
single particle mobility. Within the Drude approximation for the
mobility $T_\lambda$ is determined by the friction coefficient (or
by the viscosity for dense liquids). To check the validity of the
criterion the resulting formula is applied to non-ideal Bose gas,
to liquid helium and to $^3$He-$^4$He mixtures. A reasonable
qualitative agreement with available experiments is obtained in
all cases. In the case of the Bose gas the present approach
results also in an upper bound on the peak value of
$T_{\lambda}/T_{\lambda}^0$, where $T_{\lambda}^0$ is the critical
temperature for the ideal gas.
\end{abstract}

]

\section{Introduction}

Recently considerable attention has been attracted to the long
standing problem of how the critical temperature of the superfluid
transition depends on density (or interaction strength) in
interacting Bose systems
\cite{Lee,Matsubara,Toyoda,Stoof,Huang,Baym,Gruter,Hol,Hol2,Reppy,Wilk,Schakel,ap,Ziegler}.
If the interaction is switched off then the temperature of the
Bose-Einstein condensation is given by the well known ideal gas
formula

\begin{equation}
T_{\lambda
}^{0}=\frac{2\pi}{[\zeta(3/2)]^{2/3}}\frac{\hbar^2}{m}n^{2/3}\simeq
3.31\frac{\hbar ^{2}}{m}n^{2/3} \label{id}
\end{equation}
where $n$ is the density and $m$ is the particle mass (we use
units $\kappa _{B}=1$). This formula should be approximately valid
also for an interacting system in the limit of low density. By now
it is generally believed that if the density increases $T_{\lambda
}$ first rise with respect to the ideal gas value, but the exact
form of the correction to Eq. (\ref{id}) still remains, to some
extent, a point of controversy, despite the large number of recent
investigations
\cite{Stoof,Huang,Baym,Gruter,Hol,Hol2,Reppy,Wilk,Schakel}.

While for a weakly interacting gas there exists (at least
formally) a small parameter, proportional to the scattering
length, the problem looks even more difficult if the density is
not small. The critical temperature should finally goes down if
the density is further increased, as observed e.g. in liquid
helium, primarily due to an increase in effective mass, that is
believed to replace $m$ in Eq. (\ref{id}) for an interacting
system. Since $T_{\lambda }$ is obviously not an universal
quantity (unlike critical exponents it depends strongly on
microscopic details of the system) it seems reasonable simply to
evaluate the transition temperature numerically for a given
system, using e.g. the Path Integral Monte Carlo (PIMC) method
\cite{Ceperley,CeperleyR}, but some analytical results, if
available, would certainly lead to a better understanding. In case
of dense systems with strong interaction a lattice model may
provide a good starting point (see e.g.
Refs.\onlinecite{Matsubara,Ziegler}) though the lattice introduces
some additional commensuration effects not present in continuous
systems.

On the other hand a simple phenomenological superfluidity
criterion was proposed recently by the author \cite{ap}, which
results in an analytical expression for $T_{\lambda }$ in a dense
system. The approach is based on a Feynman picture of a bosonic
system evolving in imaginary time, when the superfluidity is
related to the appearance of permutation exchange cycles of
arbitrary large length \cite{Feynman,CeperleyR}. At high
temperatures only the identity permutation is important in the
partition function and each particle returns to its initial
position in the imaginary 'time' $\beta =1/T$ . Its root mean
square (r.m.s.) displacement $R$ is essentially the corresponding
quantum coherence length which increases when the temperature is
lowered. The superfluid transition occurs when $R$ finally reaches
the value of the order of the interparticle distance $a=n^{-1/3}$
(at lower temperatures nontrivial permutations can be no longer
ignored). This picture suggests that one can first evaluate $R(T)$
at a given temperature $T$ and then obtain an estimate for
$T_{\lambda }$ e.g. from the condition $R^2(T_\lambda)=\xi a^2$,
where $\xi$ is some numerical factor of order unity (see Sec. II
for details).

This approach is similar in spirit to the well known Lindemann
melting criterion, which states that a solid melts whenever the
r.m.s. displacement of particles from their equilibrium positions
(due to lattice vibrations) reaches a certain fraction of the
nearest-neighbor distance. It is known that in many cases the
ratio of these two quantities (the Lindemann ratio) is almost
universal thus making the criterion quite useful for qualitative
estimates (see e.g. Ref. \onlinecite{Lind}). One may hope that the
present approach which treats the superfluid transition as some
kind of 'melting in imaginary time' may also be of use at least as
a first approximation.

From the physical point of view $R$ may be regarded as a
generalization of the thermal de-Broigle wavelength $\lambda
_{T}=(2\pi \hbar ^2/mT)^{1/2}$, which is the coherence length for
non-interacting particles at a temperature $T$. Indeed, for the
ideal gas $R\sim \lambda_T$ and our criterion coincides with the
textbook argument which states that Bose-Einstein condensation
occurs whenever $\lambda_T \sim a$ (which yields $T_{\lambda}\sim
(\hbar^2/m)n^{2/3}$). If the interaction is switched on,
interparticle scattering will tend to reduce the coherence length,
so that $R$ can be much smaller than $\lambda_T$. The physics here
is essentially the same as in the decoherence due to environment
\cite{Kiefer}, since for a given particle all other particles will
now act as some dissipative environment. One may also recall that
the kinetic energy $K(T)$ of a particle in a quantum liquid is no
longer given by $\frac{3}{2}T$ but is higher due to interaction
(at low temperatures the mean kinetic energy tends to a constant
value determined by a zero-point motion), so that the particle
wavelength estimated as $\sim (\hbar ^2 /mK(T))^{1/2}$ indeed gets
smaller than $\lambda_T$.

If the interaction leads to the reduction of the coherence length
it also reduces the superfluid transition temperature (the initial
growth of $T_{\lambda }$ with interaction strength in a Bose gas
has a different origin \cite{Gruter} and will be discussed in Sec.
III). Thus e.g. the observed decrease of $T_{\lambda}$ with
pressure in helium may be understood either as a decoherence
phenomenon or as a consequence of the more intensive zero-point
motion in a pressurized system.

In Ref. \onlinecite{ap} this supefluidity criterion was applied to
molecular hydrogen H$_2$ and results in $T_{\lambda }\sim 1$ K for
the temperature of the possible superfluid transition
\cite{Ginzburg,Maris,Ceperley1,Vorob'ev}. This estimate is
significantly lower than the very first one, based on the ideal
gas formula (\ref{id}), $6\div 8$ K \cite{Ginzburg} (though
consistent with later estimates \cite{Maris} and PIMC results for
H$_2$ clusters \cite{Ceperley2} and films \cite{Ceperley1}). This
result was obtained, however, by assuming that the 'Lindemann'
parameter $\xi$ is universal, i.e. the same as in case of liquid
helium. Such an assumption may cause some doubt and its further
analysis is certainly in order. The r.m.s. displacement $R$ that
enters the superfluidity criterion was evaluated in Ref.
\onlinecite{ap} using the Caldeira-Leggett model \cite{Caldeira}
which also needs some justification in our case.

In this paper we study this problems in more detail in order to
better understand the accuracy and reliability of the criterion
proposed. In Sec. II we review some general aspects of the theory.
The general formula will be derived, which relates  $R^2$ to the
mobility of an external particle in the liquid. We will show that
the Caldeira-Leggett model approach is equivalent in our case to
the Drude approximation for the mobility. Within the Drude
approximation the mobility is determined entirely by the collision
rate $\gamma $ and it appears possible to derive a general formula
for the superfluid transition temperature which is an universal
function of $\hbar \gamma /T_{\lambda }^{0}$. If this ratio is
much larger than unity the superfluidity is suppressed.

Next we apply the formula obtained to different superfluid systems
and compare the results to available experimental data. In Sec.
III a weakly interacting Bose gas is considered. Here we analyze
the ratio of $T_{\lambda }/T_{\lambda }^{0}$ as a function of
density and compare the result to the known PIMC calculation
\cite{Gruter}. Sec. IV is devoted to liquid helium. Contrary to
the previous work \cite{ap} where $\gamma $ was related to the
mean kinetic energy of a helium atom, the transition temperature
is expressed here as a function of the shear viscosity in the
normal state (at $T$ close to $T_{\lambda}$). We also check the
validity of the Drude approximation and find out that for liquid
helium this is quite accurate if exchange effects are neglected.
Finally, in Sec. V we extend the present approach to
$^{3}$He-$^{4}$He mixtures. With no additional fitting it appears
possible to describe qualitatively the behaviour of $T_{\lambda }$
as a function of both the total density and the concentration of
$^{3}$He atoms.

\section{General formalism}

According to the Feynman's path integral approach \cite{Feynman}
the partition function $Z$ for a system of $N$ interacting bosons
can be represented as a sum over all possible permutations of
these particles
\begin{equation}
Z=\frac{1}{N!}\sum_{P}\int \prod_{i}d{\bf r}_{i}\int
\prod_{i}{\cal D}{\bf r} _{i}(\tau )\exp (-\frac{1}{\hbar }S)\;,
\label{z}
\end{equation}
\[
S=\int_{0}^{\hbar \beta }\lbrack \sum_{i}\frac{m{\bf
\dot{r}}_{i}^{2}}{2} +\sum_{i<j}V({\bf r}_{i}-{\bf r}_{j})\rbrack
d\tau
\]
where $V({\bf r}_{i}-{\bf r}_{j})$ is the interparticle
interaction potential and $\beta =1/T$. The integration in Eq.
(\ref{z}) is over all paths with ${\bf r}_{i}(0)={\bf r}_{i},$
${\bf r}_{i}(\hbar \beta )=P{\bf r}_{i}$, where $P$ is some
permutation of $N$ particles.

At high temperatures only the identity permutation contributes to
the sum, but as $\beta $ increases, other permutations finally
also become important and the system becomes superfluid. In this
picture the superfluid transition is related to the appearance of
long cyclic permutations (see e.g. Ref. \onlinecite{CeperleyR} for
details). This means that the critical temperature should be
roughly of the same order as a temperature at which nontrivial
permutations can no longer be neglected. Then we may try to
estimate $T_{\lambda}$ from a criterion \cite{ap}
\begin{equation}
R^{2}\equiv {\frac{1}{\hbar \beta }}\int_{0}^{\hbar \beta }\langle \lbrack
{\bf r}{(\tau )}-{\bf r}{(0)}\rbrack ^{2}\rangle d\tau =\xi a^{2},
\label{cr}
\end{equation}
where ${\bf r}(\tau)$ is a path of some arbitrary particle in the
system of {\em distinguishable} particles and $\xi $ is a
numerical factor, to be determined later. Average in this formula
means the functional integration over all paths of the chosen
particle ${\bf r}{(\tau )}$ with ${\bf r}{(0)=}{\bf r}{(\beta )}$

\begin{equation}
\langle \ldots \rangle ={\frac{1}{Z}}\int {\cal D}{\bf r}(\tau )\langle
\ldots \rangle \exp \left( -\frac{1}{\hbar }\int_{0}^{\hbar \beta }d\tau
\frac{m\dot{{\bf r}}^{2}}{2}-W\{{\bf r}(\tau )\}\right)
\end{equation}
where the effective action $W\{{\bf r}(\tau )\}$ arises from the
integration over other $N-1$ particles ($i=1,2\ldots N-1$) with
${\bf r}_{i}(0)={\bf r}_{i}(\beta )$
\begin{equation}
\exp \left( -W\{{\bf r}\}\right) =\int \prod_{i} {\cal D}{\bf
r}_{i}(\tau )\exp (-\frac{1}{\hbar }S^{\prime }\{{\bf r},{\bf r}
_{i}\})\;,
\label{W}
\end{equation}
\[
S^{\prime }\{{\bf r},{\bf r}_{i}\}=\int_{0}^{\hbar \beta }\lbrack
\sum_{i} \frac{m{\bf \dot{r}}_{i}^{2}}{2}+\sum_{i}V({\bf r}-{\bf
r}_{i})+\sum_{i<j}V({\bf r}_{i}-{\bf r}_{j})\rbrack d\tau
\]

The condition (\ref{cr}) merely states that near the transition
the mean displacement (in imaginary time) of a given particle from
its initial position becomes comparable to the interparticle
spacing. This is similar in spirit to the well known Lindemann
melting criterion. For this reason we shall some times refer to
Eq. (\ref{cr}) as the 'Lindemann-like' criterion and also call
$\xi$ the 'Lindemann ratio'. We shall assume here that $\xi$ is
roughly a constant, but, certainly, possible weak dependence of
$\xi$ on system parameters (e.g. on density) cannot be excluded
and  may be important in some cases (see e.g. Sec. III).

In the ideal gas the left hand side of Eq. (\ref{cr}) is
essentially the square of the de Broglie thermal wavelength
$\lambda _{T}^{2}=2\pi \hbar ^{2}/mT$, and our criterion looks
like $\lambda_{T} \sim a$, but interactions will tend to reduce
$R^{2}$ (this was observed e.g. in Ref. \onlinecite {Cleveland}).
One may think of this reduction as arising from the decoherence
due to interaction with environment. Neighboring particles which
scatter from the one we are looking at 'measure', in a sense, its
position thus reducing its quantum uncertainty in coordinate space
(compare e.g. with Ref.\onlinecite{Kiefer}).

Estimating $R^{2}$ in a system of interacting particles is still a
very complicated problem, even if exchanges are neglected. One has
first to calculate somehow the effective action $W\{{\bf r}\}$
given by Eq. (\ref{W}). However, since we are interested only in
the simple average (\ref{cr}) we can avoid the evaluation of
$W\{{\bf r}\}$ and relate this average to the mobility of an
external particle.

If we introduce a Fourier transform of the path ${\bf r}{(\tau )}$
according to
\[
{\bf r}(\tau )=\sum_{n=-\infty }^{+\infty }{\bf r}
_{n}e^{-i\omega _{n}\tau }\;,
\]
where $\omega _{n}=2\pi n/\hbar \beta $, $n=0,\pm 1,\pm 2,\ldots $
then in three dimensions the mean square displacement defined by
Eq. (\ref{cr}) can be written as
\begin{equation}
R^{2}=12\sum_{n=1}^{+\infty }\langle
|{\bf r} _{n}|^{2}\rangle
\end{equation}

Now let us write the expression for the mobility in the Matsubara
representation $\mu _{M}(\omega _{n})$. For this purpose we
introduce an external force ${\bf f}(\tau )={\bf f}\exp (-i\omega
_{n}\tau )$ acting on a given particle and a perturbation term
\begin{equation}
\delta S=-\int_{0}^{\hbar \beta }{\bf r}(\tau ){\bf f}(\tau )d\tau
\end{equation}
in the action. The force will induce the average velocity $\langle
{\bf v}(\tau )\rangle={\bf v}_n\exp (-i\omega _{n}\tau )$ and the
mobility $\mu _{M}(\omega _{n})$ is defined according to
\begin{equation}
{\bf v}_{n}=\mu _{M}(\omega _{n}){\bf f} .
\end{equation}
Since
$\langle {\bf v}(\tau )\rangle =i\langle {\bf \dot{r}} (\tau
)\rangle $, to the first order in ${\bf f}$ we obtain
\begin{equation}
\langle {\bf v}(\tau )\rangle =\frac {i}{\hbar}\int_{0}^{\hbar
\beta }\langle {\bf \dot{r}} (\tau ){\bf r}(\tau ^{\prime
})\rangle {\bf f}(\tau ^{\prime })d\tau ^{\prime },
\end{equation}
and hence
\begin{equation}
\mu _{M}(\omega _{n})=\omega _{n}\beta \langle
|{\bf r} _{n}|^{2}\rangle .
\end{equation}
This means that the
average we are interested in can be expressed as follows
\begin{equation}
R^{2}=\frac{12 }{\beta }\sum_{n=1}^{+\infty }\frac{\mu _{M}(\omega
_{n}) }{\omega _{n}} .  \label{rm}
\end{equation}
The Matsubara mobility is related to the mobility at real frequency $\mu
(\omega )$ by the usual equation
\[
\mu _{M}(\omega _{n})=\mu (i|\omega _{n}|).
\]

Since $\mu (\omega )$ is a response function and is analytical in
the upper half-plane of the complex frequency one can easily
verify that
\begin{equation}
\mu (i\omega _{n})=\frac{1}{\pi }\int_{-\infty }^{+\infty }d\omega
\frac{ \omega _{n} {\rm Re} \mu (\omega )}{\omega _{n}^{2}+\omega
^{2}}
\end{equation}
for positive $\omega _{n}$. Substituting this
expression in Eq. (\ref{rm}) and carrying out summation over $n$
we finally obtain
\begin{equation}
R^{2}=\frac{6\hbar }{\pi }\int_{0}^{+\infty }\frac{d\omega
}{\omega } {\rm Re} \mu (\omega )\left[ {\rm \coth }(\frac{\hbar
\omega }{2T})-\frac{2T}{\hbar \omega }\right] .  \label{rf}
\end{equation}

Thus if we know $\mu (\omega )$ from some theory or experiment, we
can evaluate the mean square displacement (\ref{rf}) which enters
in our superfluidity criterion.

Important asymptotics can be obtained, however, without knowing
the exact expression for $\mu (\omega )$. We shall assume only
that the static mobility $\mu =\mu (0)$ is finite and introduce
the scattering rate $\gamma =1/(m\mu )$. First of all , if the
interaction is switched off, then $R^{2}\rightarrow \hbar
^{2}\beta /2m$. Then the criterion (\ref{cr}) results in
$T_{\lambda }=(1/2\xi )\hbar ^{2}n^{2/3}/m$. Since this is the
ideal gas formula for the critical temperature (\ref{id}) we
conclude that $1/2\xi \simeq 3.31$, i.e.

\begin{equation}
\xi \simeq 0.15
\end{equation}

If however the interaction is so strong that $\hbar \gamma \gg T$
then the leading low temperature contribution to $R^{2}$ will be
logarithmic in $\hbar \gamma /T$
\begin{equation}
R^{2}\simeq \frac{6}{\pi }\frac{\hbar }{m\gamma }\ln (\frac{\hbar
\gamma }{2\pi T})+{\rm const}  \label{log}
\end{equation}

It is clear from (\ref{rf}) that this logarithmic behavior is
actually independent of the form of $\mu (\omega )$ provided the
static mobility is finite i.e. $\mu (0)\neq 0$, and the prefactor
is determined entirely by $\mu =1/(m\gamma )$. The second term in
(\ref{log}) depends on the exact form of the function $\mu (\omega
)$.

It is important, that though we consider low temperatures, we
always deal here with nondegenerate liquid where the mobility (and
hence $\gamma $) is approximately independent of temperature (see
Sec. III). Below the $\lambda $ point the friction of an external
particle is determined by collisions with quasiparticles (phonons
and rotons) and will tend to zero as $T\rightarrow 0$. For this
reason it is more adequate to describe the system below the
$\lambda$ point in terms of an effective mass $m^{\ast}$. Our
present approach is in fact complementary to the one based on the
notion of effective mass, which results in $T_{\lambda}\sim
(\hbar^2/m^{\ast})n^{2/3}$. The temperature dependence of $R^2$ in
the normal system above $T_{\lambda}$ (with all exchange effects
switched off) can hardly be described by a simple renormalization
of the particle mass, and the incoherent diffusion may be more
adequate a model for the particle motion (see also Sec. IV).
Though the effective mass approach results in the apparently
simple expression for $T_{\lambda}$ it is not really simpler,
since there are no analytical closed expressions for $m^{\ast}$.

Using Eq. (\ref{log}) we can obtain from Eq. (\ref{cr}) a very
simple formula \cite{ap}
\begin{equation}
T_{\lambda }=\alpha \frac{\hbar \gamma }{2\pi }\exp \left(-\xi
\frac{\pi }{6} \frac{\hbar \gamma }{T_{0}}\right)\;,\qquad
T_{0}=\frac{\hbar ^{2}}{m}n^{2/3} \label{exp}
\end{equation}
where $\alpha $ is some unknown factor, dependent on the high
frequency behavior of $\mu (\omega )$, and $T_0$ is essentially
the critical temperature for the ideal gas (it differs from Eq.
(\ref{id}) only by a numerical factor).

Thus the temperature of the superfluid transition crucially
depends on the ratio $\hbar \gamma /T_{0}$ which is a function of
density and interaction strength. At $\hbar \gamma \ll T_{0}$ the
ideal gas formula (\ref{id}) is valid while in the opposite limit
the critical temperature of the $\lambda $ transition is
exponentially small due to the decoherence phenomenon. Since
$\gamma $ should normally increase with density (for systems with
strong repulsion between particles at small distances), the
formula obtained qualitatively explains the suppression of
$T_{\lambda }$ in sufficiently dense systems.

Now we introduce a simple model for $\mu (\omega )$. At high
frequencies (higher than the collision rate) the mobility is the
same as for a free particle, i.e. $\mu (\omega )\sim i/(m\omega
)$. Then we may try to use, as a first approximation, the
interpolation formula for the frequency dependent mobility in the
Drude form
\begin{equation}
\mu (\omega )=\frac{1}{\lambda -im\omega }  \label{dr}
\end{equation}
where $\lambda $ is the friction coefficient which determines the
static mobility $\mu =\mu (0)=1/\lambda $. Frequency dependence of
this kind is valid for the mobility of a heavy particle, moving
with friction according to the Langevin equation. Variational
calculations of Ref. \onlinecite{Rosch} suggest that the Drude
formula may be adequate for a light particle as well. In fact this
interpolation formula is reasonable when there is only one
frequency scale in a problem. Strictly speaking the friction
coefficient should be also frequency dependent, i.e. $\lambda
=\lambda (\omega )$, but in many cases it is possible to neglect
this dispersion since ${\rm Re} \mu (\omega )$ will be already
very small when the dispersion in $\lambda $ becomes significant.
Now, substituting Eq. (\ref{dr}) in Eq. (\ref{rf}) we obtain
\begin{equation}
R^{2}=\frac{6}{\pi }\frac{\hbar }{m\gamma }\left[ C+\psi \left(
1+\frac{\hbar \gamma \beta }{2\pi }\right) \right]  \label{r2}
\end{equation}
where $\gamma =\lambda /m,$ $\psi (x)$ is the digamma function and
$C\simeq 0.577\ldots $ is the Euler's constant. This very
expression was used previously \cite{ap} when the motion of a
particle was described by the Caldeira-Leggett model with ohmic
dissipation. Note that this formula (with the suitable choice of
$\gamma$) may be used also to describe the suppression of
superfluidity by an external disordered potential (this will be
discussed elsewhere).

The final result for the transition temperature in case of $\mu
(\omega )$ given by Eq. (\ref{dr}) may be written as
\begin{equation}
T_{\lambda }=\frac{\hbar \gamma }{2\pi }F\left(\xi \frac{\pi
}{6}\frac{\hbar \gamma }{T_{0}}\right),\qquad \gamma =\lambda /m
\label{td}
\end{equation}
where the function $F(x)$ is implicitly defined by the equation
\begin{equation}
x=C+\psi (1+1/F(x)).  \label{eq}
\end{equation}
From Eq. (\ref{eq}) one can obtain $F(x)\simeq (6/\pi ^{2})1/x$ at
$x\rightarrow 0$ while $F(x)$ becomes exponentially small
$F(x)\simeq \exp (C-x)$ at $x\rightarrow \infty $. In the
intermediate region $x\sim 1$ this function is well approximated
by the interpolation formula $F(x)\simeq 1.6\exp (-0.96x)/(1-\exp
(-0.96x))$.

Unfortunately, even for such a simple choice of $\mu (\omega )$
there seems to be no opportunity to calculate the value of $\gamma
$ analytically for a dense liquid with a strong interaction
between particles. For this reason one has to use experimental
results as well as approximations with some adjustable parameters.
One can also try to use the fluctuation-dissipation theorem for
the velocity fluctuations and write the following expression for
the kinetic energy
\begin{equation}
K(T)=\frac{3}{2}m\int_{0}^{\infty }\frac{d\omega }{\pi }\hbar
\omega {\rm Re}\mu (\omega ){\rm \coth }(\frac{\hbar \omega }{2T})
\label{K}
\end{equation}
If we put $T=0$ then
\begin{equation}
K=\frac{3}{2}m\int_{0}^{\infty }d\omega \hbar \omega {\rm Re} \mu
(\omega )  \label{k0}
\end{equation}
where $K=K(0)$. This integral is known to diverge logarithmically for the
Drude mobility at high frequencies. If we assume that for true $\mu (\omega
) $ there is only one characteristic frequency scale $\gamma $ then
obviously
\[
K\sim m\hbar \gamma ^{2}/\lambda \sim \hbar \gamma
\]
This approximation was used in the previous paper \cite{ap}, and
it has some advantage, because the kinetic energy $K$ (essentially
the zero-point energy) can be easily evaluated for various systems
\cite{Dyugaev}. In what follows, however, we shall try to make a
more direct estimate of the damping parameter $\gamma $.

\section{Weakly interacting Bose gas}

Before we turn to dense liquids consider first a non-ideal Bose
gas. In the gas the transition temperature is close to Eq.
(\ref{id}) and we are interested in corrections to $T_{\lambda
}^{0}$ at non-zero density. To estimate these corrections consider
a classical gas of hard spheres of diameter $d$. Then the
scattering rate is given by
\begin{equation} \label{scat}
\gamma \sim v/l\sim \pi nd^{2}v
\end{equation}
where $l\sim 1/(n\pi d^{2})$ is the mean free path and $v\sim (T/m)^{1/2}$
is the mean velocity. Since $T\sim T_{\lambda }^{0}\sim \hbar ^{2}n^{2/3}/m$

\begin{equation}
\gamma \sim \frac{\hbar}{m} d^{2}n^{4/3}  \label{fr_gas}
\end{equation}
If we introduce a small parameter $z=(nd^{3})^{1/3}$ then we see
that the ratio $\hbar \gamma /T_{\lambda}^0\sim z^{2}$ is small at
low densities and one can expand the formula (\ref{r2}) for $R^2$
in $\hbar \gamma /T$. Retaining only the first correction in
$\gamma$ we find from Eq. (\ref{cr})
\begin{equation}
T_{\lambda }\simeq \frac{1}{2\xi }\frac{\hbar^2}{m}n^{2/3}\left(
1-\frac{3\zeta (3)}{\pi ^{3}} \frac{\hbar \gamma }{T_{\lambda
}^{0}}\right)
\end{equation}
The correction due to decoherence is negative and leads to the
suppression of $T_{\lambda }$. It is generally accepted, however,
that the first correction to Eq. (\ref{id}) should be positive
\cite{Stoof,Huang,Baym,Gruter,Hol,Hol2,Schakel}. The only way to
account for this increase of $T_{\lambda }$ in the present
approach is to assume that the parameter $\xi $ also slightly
depends on density and gets smaller as the density is increased
i.e.
\begin{equation}
\xi =\xi (n)\simeq \xi _{0}(1-\delta )  \label{xi}
\end{equation}
where $\xi _{0}\simeq 0.15$ and the small density dependent
correction $\delta $ will be specified later. The physical reason
for such a dependence is the same as in Ref. \onlinecite{Gruter}:
particles in a more dense system  tend to be more homogeneously
distributed through the whole volume, hence it is easier for a
particle to find a neighbor for exchange. Then we can write
\begin{equation}
\frac{T_{\lambda }}{T_{\lambda }^{0}}\simeq 1+\delta -\frac{3\zeta
(3)}{ \pi ^{3}}\frac{\hbar \gamma }{T_{\lambda }^{0}}
\label{t_gas}
\end{equation}

Now we need some more accurate estimate of $\gamma$.  For a
nondegenerate system of hard spheres one can evaluate the friction
coefficient $\lambda=m\gamma$ from the Bolzmann equation and in
the low density limit
\begin{equation}
\lambda =\frac{32}{3}nd^{2}(\pi mT)^{1/2}
\end{equation}
(see e.g. Ref. \onlinecite{Crox}). The additional factor of $4$ is
introduced here because in our case, contrary to the pure
classical one, the wavelength of the particle is larger then $d$
and the $s$-wave scattering cross section is known to be $4\pi
d^{2}$ rather than $\pi d^2$ as for a classical sphere.

Hence near the transition, when $T\simeq T_{\lambda}^0$
\begin{equation}
\hbar \gamma \simeq \frac{32\sqrt{2}\pi}{3[\zeta
(3/2)]^{1/3}}\frac{\hbar ^{2}}{m} d^{2}n^{4/3}  \label{gamma}
\end{equation}

Most of recent calculations suggest, that at low density the first
positive correction in Eq. (\ref{t_gas}) should be of the first
order in $d$ \cite{Stoof,Baym,Gruter,Hol,Hol2}, so in this limit
we can take $\delta \sim z$. At higher density the Lindemann ratio
is likely to reach some finite limit close to $\xi \simeq 0.12$.
\cite{ap} Then we may try a simple interpolation formula
\begin{equation}\label{int}
\delta =\frac{a_{1}z}{1+a_{2}z}
\end{equation}

Then the final expression for the transition temperature can be written as
\begin{equation}
T_{\lambda }/T_{\lambda }^{0}\simeq 1+\frac{a_{1}z}{1+a_{2}z}
-a_{3}z^{2},\qquad z=(nd^{3})^{1/3}
\label{tf_gas}
\end{equation}
where $a_{3}\simeq 1.2$ from Eqs.(\ref{t_gas}) and (\ref{gamma}),
while $a_{1}$ and $a_{2}$ should be taken from some microscopic
theory or experiment. The dependence of $T_{\lambda }/T_{\lambda
}^{0}$ on density given by Eq. (\ref {tf_gas}) is shown in Fig. 1
 for $a_{1}\simeq 2.3$ as obtained e.g. in recent Monte Carlo
study \cite{Hol} and for $a_{2}\simeq 9.9$. This value of $a_2$
was fixed by the requirement that at $z\simeq 0.61$ (which
approximately corresponds to helium density, when $nd^3\simeq
0.23$) $\xi $ should be close to $0.12$ i.e. $\delta \simeq 0.2$.
The squares represent the PIMC data \cite{Gruter} for a quantum
system of hard spheres. Though we have taken rather large initial
slope $a_{1}\simeq 2.3$, which is almost an order of magnitude
larger than reported in Ref. \onlinecite{Gruter}, the resulting
curve does not deviates much from the PIMC data (the largest
deviation is less than ten percent) and reveals a qualitatively
similar behaviour.

\begin{figure}[htp]
\epsfxsize=3.375in \centerline{\epsffile{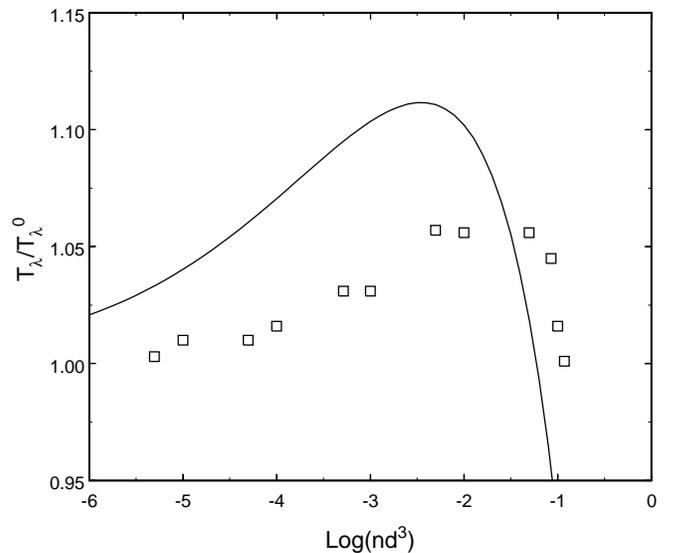}}

\caption{Critical temperature $T_\lambda$ of an interacting Bose
gas versus density from Eq. (\protect\ref{tf_gas}) with $a_1=2.3$,
$a_2=9.9$ and $a_3=1.2$ ($T_\lambda ^0$ is the critical
temperature of the ideal gas, $d$ is the hard-sphere diameter).
For comparison, PIMC simulation results \protect\cite{Gruter} are
also included.} \label{fig1}
\end{figure}

What is perhaps more interesting, explaining the initial rise of
$T_{\lambda }$ with density by a change in the Lindemann ratio
$\xi $ places an upper bound on the peak value of $T_{\lambda
}/T_{\lambda }^{0}$. Indeed, if the parameter $\xi $ changes from
$\xi _{0}\simeq 0.15$ to something like $\xi \sim 0.12$ at high
densities then the maximum value of $\delta $ in Eq. (\ref{t_gas})
is $0.2$ and hence $T_{\lambda }/T_{\lambda }^{0}$ should not
exceed $1.2$. This result is consistent with the PIMC calculations
\cite{Gruter} but is incompatible with recent experiment in
$^{4}$He-Vycor system \cite{Reppy} where the peak value of
$T_{\lambda }/T_{\lambda }^{0}$ was found to be almost $3.5$ times
larger. This disagreement seems to be not yet explained and one
may suspect that we do not fully understand all possible effects
of a disordered porous substrate in Vycor samples on
$T_{\lambda}$.

\section{Liquid helium}

Let us see first what an elementary kinetic approach can suggest
for $\mu (\omega )$. It is known from experiment \cite{Wood} that
above the superfluid transition the momentum distribution of
helium atoms is approximately of a Gaussian type with the mean
kinetic energy $K\sim 15$ K. If we used a Boltzman equation for a
classical gas of hard spheres of diameter $d$ at effective
temperature $T^{\ast }=\frac{2}{3}K$ then in the relaxation time
approximation we obviously would obtain the frequency dependent
mobility in the Drude form (\ref{dr}) with the inverse relaxation
time $\gamma \sim v/l\sim \pi nd^{2}v$ as in Eq. (\ref{scat})
where now $v\sim (K/m)^{1/2}$. At the same time one has
\[
\lambda \sim m\gamma \sim \pi nd^{2}mv
\]
for the friction coefficient and
\begin{equation}
\eta \sim mvnl\sim (1/\pi )mv/d^{2}  \label{vis}
\end{equation}
for shear viscosity. From these formulas we get
\begin{equation}
\lambda \sim \pi ^{2}(nd^{3})\eta d  \label{lkin}
\end{equation}
Extrapolating this result to a dense system with $nd^{3}\sim 1$ we
see that the friction coefficient in a liquid is in fact
proportional to the viscosity. Note that the estimate (\ref{vis})
is quite reasonable for liquid helium above the $\lambda $ point
\cite{Taylor}. Indeed, the measured viscosity above $T_{\lambda }$
is almost independent of temperature and is equal to $\eta \simeq
3.5\cdot 10^{-5}$ poise (at s.v.p. and at $T=3.2$ K) while the
more accurate version of Eq. (\ref{vis}) $\eta \simeq
0.18(mT^{\ast })^{1/2}/d^{2}$  gives $\eta \simeq 3.6\cdot
10^{-5}$ poise at $d\simeq 2.2$ \AA .

Practically the same result as Eq. (\ref{lkin}) arises also from
the hydrodynamic approach. For a particle of a large radius ${\cal
R}\gg a$ in a liquid the friction coefficient $\lambda $ from the
Stokes' formula is $6\pi \eta {\cal R}$. It is clear, however,
that for a particle of the liquid, which is not large as compared
to other particles, this formula should be modified. Long ago it
was argued (see e.g. Ref. \onlinecite{Herzog} and references
therein) that it this case one should rather take $\lambda =4\pi
\eta {\cal R}$, as for an empty bubble in a liquid. One may
suppose in this connection that a particle in a liquid is
surrounded by a 'correlation hole', so that one may view the
particle as associated with a 'bubble' of radius ${\cal R}\sim
n^{-1/3}$. Then we have an estimate
\begin{equation}
\lambda =4\pi \eta bn^{-1/3}  \label{l}
\end{equation}
where we have put ${\cal R}=bn^{-1/3}$ and $b$ is some numerical
factor of order unity. Certainly, this estimate makes sense only
above the $\lambda $ transition and for sufficiently dense liquid.
When $nd^{3}\sim 1$, estimates (\ref{lkin}) and (\ref{l}) lead to
essentially the same results. In what follows we shall use Eq.
(\ref{l}) since this will result in a slightly more transparent
final formulas.

Now, making use of Eq. (\ref{l}) we can rewrite the general
formula (\ref{td}) for the superfluid transition temperature  in
the form
\begin{equation}
T_{\lambda }=A\frac{\hbar ^{2}}{m}n^{2/3}\frac{\eta }{\eta
_{0}}F\left(B\frac{\eta }{\eta _{0}}\right),\qquad \eta _{0}=\hbar
n \label{h}
\end{equation}
where the function $F(x)$ is defined by Eq.(\ref{eq}) and

\begin{equation}
A=2b,\qquad B=\frac{2}{3}\pi ^{2}b\xi .  \label{ab}
\end{equation}

We see now that the critical temperature in this approach is
determined by the shear viscosity $\eta $, and $T_{\lambda }$ will
be smaller than the ideal gas value if the viscosity is
considerably larger than $\eta _{0}=\hbar n$, which may be called
a 'quantum' of viscosity. For typical densities of liquid helium
one has $\eta _{0}\simeq 2\cdot 10^{-5}$ poise, while the
viscosity above the superfluid transition is $\eta \simeq 3.5\cdot
10^{-5}$ poise. Thus in the case of helium $\eta $ is only
slightly larger than $\eta _{0} $, and hence the temperature of
the $\lambda $ transition should be close to that of the ideal
gas, as it is experimentally. In sufficiently viscous Bose liquids
the critical temperature of the possible superfluid transition
should be suppressed and they will crystallize on cooling.

The viscosity in helium slightly above $T_{\lambda }$ is almost
independent of temperature and increases with density
approximately by a factor of two when the pressure is raised from
s.v.p. to $24$ atm. We take experimental data for the pressure
dependent viscosity at $T=3.2$ K from Ref. \onlinecite{Goodwin}
and turn it into $\eta (n)$ dependence using the known equation of
state at the same temperature. The density dependence of the ratio
$\eta /\eta _{0}$ is approximated then by the second order
polynomial
\begin{equation}
\eta /\eta _{0}\simeq \sum_{l=0}^{2}c_{l}(n^{\ast })^{l}  \label{pol}
\end{equation}
where $n^{\ast }=nr_{0}^{3}$ ($r_{0}\simeq 2.556$ \AA ) is the
reduced density and the coefficients are $c_{0}\simeq 2.965$,
$c_{1}\simeq -15.701$ and $c_{2}\simeq 34.027$. Then we fit Eq.
(\ref{h}) to the experimental data for $T_{\lambda }$ in helium
(actually to the Kierstead empirical equation for the $\lambda $
line \cite{Kierstead}) and obtain $A\simeq $ $1.39$ and $B\simeq
0.58$, which correspond to quite reasonable values
\begin{equation}
b\simeq 0.7,\qquad \xi \simeq 0.13  \label{fit}
\end{equation}
The resulting curve for the transition temperature is shown in
Fig. 2 (solid line). We also extrapolate this curve to lower
densities, corresponding to a metastable liquid, though we do not
expect our hydrodynamic formula (\ref{l}) to be valid at $n\sim
0.2\div 0.3$. At low densities the friction coefficient should be
smaller than Eq. (\ref{l}) (see e.g. Eq. (\ref{lkin}))and hence
the true value of $T_{\lambda}$ should be higher. Also on this
figure the condensation temperature for the ideal gas is shown
(dashed line) along with the experimental data for liquid helium
(squares) and recent PIMC results for metastable helium at
negative pressure \cite{Bauer} (circles). Note that the points at
negative pressure were not used in the fit.

\begin{figure}[htp]
\epsfxsize=3.375in \centerline{\epsffile{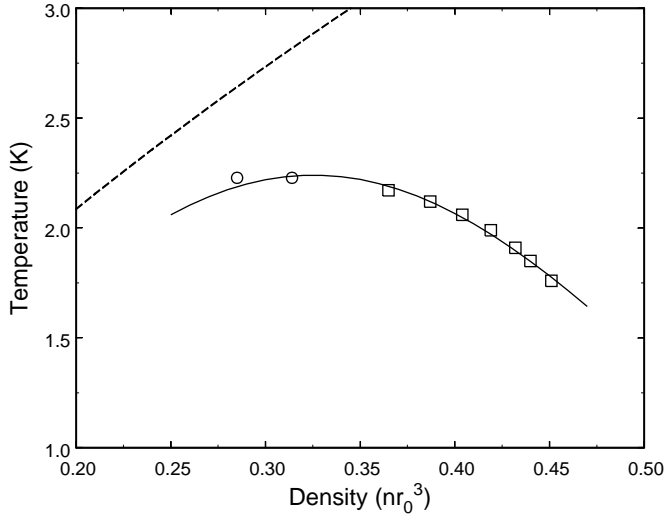}}

\caption{ Temperature of the superfluid transition versus the
reduced density $n^{\ast}=nr_{0}^{3}$ ($r_{0}=2.556$ \AA). Solid
line is the theory (\protect\ref{h}) at $A=1.39$, $B=0.58$,
squares denotes experimental data for helium, dashed line
corresponds to the ideal gas. Recent simulation results for helium
at negative pressure \protect\cite{Bauer} are also included
(circles).}

\label{fig2}
\end{figure}

The value of the Lindemann ratio $\xi $ from Eq. (\ref{fit}) is
quite close to that of the ideal gas $\xi _{0}\simeq 0.15$ and
also to $\xi \simeq 0.12$ obtained in Ref. \onlinecite{ap} by
taking $\hbar \gamma \sim K$. If we now evaluate the value of the
damping parameter $\gamma =\lambda /m$ then at $b\simeq 0.7$ we
have from Eq. (\ref{l}) $\hbar \gamma \simeq 13$ K, which is
indeed close to the zero point kinetic energy $K\simeq 15$ K.

In the case of helium it appears possible to check also the
validity of the Drude approximation (\ref{dr}) which results in
Eq. (\ref{r2}) for the mean square displacement in imaginary time.
For this purpose let us consider a more general quantity
\[
R^{2}(\tau )=\langle \lbrack {\bf r}{(\tau )}-{\bf r}{(0)}\rbrack
^{2}\rangle
\]
The $R^{2}$ discussed above is just the average of $R^{2}(\tau )$
over imaginary time. Proceeding as in Sec. II and taking the
mobility in the Drude form (\ref{dr}) we obtain
\begin{equation}
R^{2}(\tau )=\frac{24}{m\beta }\sum_{n=1}^{+\infty
}\frac{1}{\omega _{n}^{2}+\gamma \omega _{n}}\sin ^{2}\left(
\frac{\omega _{n}\tau }{2}\right) \label{r2t}
\end{equation}
Note that for a free particle one would have
\begin{equation}
R^{2}(\tau )=(3/m\beta )\tau (\hbar \beta -\tau )
\end{equation}
At small $\tau $ we can replace the sum in Eq. (\ref{r2t}) by the
integral to obtain $R^{2}(\tau )\simeq (3\hbar /m)\tau $ which is
independent of $\gamma $ since on a very short time scale the
particle moves without collisions. In the general case the sum can
be calculated numerically and the result for $\beta =2$
K$^{-1}$and $\hbar \gamma \simeq 13 $ K (as obtained from the
fitting of $T_\lambda$) is shown by the solid line in Fig. 3. Also
the PIMC calculation results are shown as presented in Ref.
\onlinecite{CeperleyR}, where the same quantity was evaluated for
bosons (triangles) and for distinguishable particles (squares).

\begin{figure}[htp]
\epsfxsize=3.375in \centerline{\epsffile{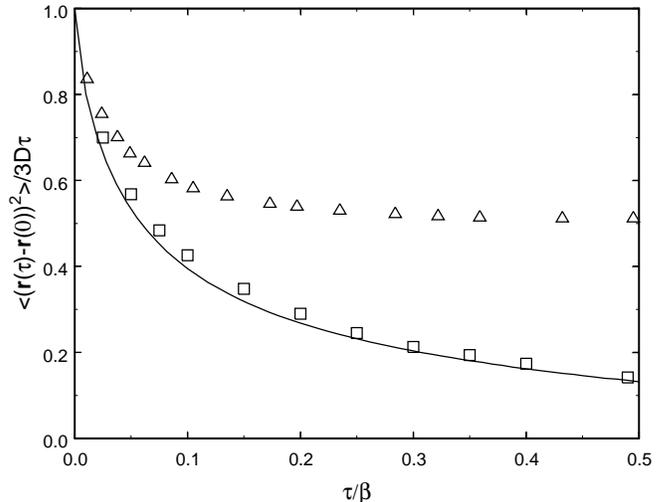}}

\caption{The diffusion of $^4$He atoms as a function of imaginary
time $\tau$ for $T=0.5$ K ($D=\hbar /m$ and $\beta=1/T$). The
solid line results from Eq. (\protect\ref{r2t}) at $\hbar \gamma
=13 K$. For comparison, PIMC simulation results
\protect\cite{CeperleyR} are also included, both for bosons
(triangles) and distinguishable particles (squares).}

\label{fig3}
\end{figure}

Is is clear from Fig. 3 that the Drude approximation well account
for the depression of $R^{2}(\tau )$ due to the interaction in the
nondegenerate system. Since by our definition $R^{2}$ should be
calculated for the system with all exchange effects neglected (see
Sec. II) we conclude that the Drude approximation (and hence the
Caldeira-Leggett model) may be quite sufficient for our
superfluidity criterion.

\section{$^{3}$He-$^{4}$He mixtures}

To check the universality of the Lindemann-like criterion it is
necessary to consider some other superfluid systems. A natural
choice to start with is the case of $^{3}$He-$^{4}$He mixtures.
One may try to repeat calculations from the previous section and
express $T_\lambda$ in terms of viscosity. Unfortunately there
seems to be no such viscosity measurements  for a wide range of
pressures as that of Goodwin \cite{Goodwin} for pure $^{4}$He.
Though there exists a recent investigation \cite{Wang} of $\eta $
in mixtures it deals only with pressures near the saturated vapor
pressure. Above $T_\lambda$ (at $T \sim 3$ K) the viscosity at
s.v.p. is almost independent of temperature and gets smaller if
the concentration $x$ of $^{3}$He atoms increases.

At small $x$ this effect may be attributed simply to the decrease
of the total density. Indeed, adding $^{3}$He atoms to liquid
$^{4}$He leads to some reduction of the mean density (due to
lighter mass of $^{3}$He atoms they have larger zero-point energy
and are harder to localize) and hence to a lower viscosity.  But
at a fixed density the local environment of a given $^{4}$He atom
remains practically the same as for $x=0$ since interactions
between atoms do not depend on the isotope mass. Hence the
viscosity and the mean square displacement $R^{2}$ in dilute
mixtures (small $x$) are likely to be the same as in pure $^{4}$He
at the same density.

However, it is the mean distance between Bose particles that
enters in the r.h.s. of the criterion (\ref{cr}). This distance is
now larger by the factor $(1-x)^{-1/3}$ due to the presence of
additional $^{3}$He atoms which do not participate in
superfluidity, so the criterion should read
\begin{equation}
R^{2}=\xi (1-x)^{-2/3}a^{2}
\end{equation}
where $a=n^{-1/3}$ is determined by the total mean density.

Now assuming that at a {\em fixed density} $R^{2}$ is indeed the
same as in pure $^{4}$He, and proceeding as in the previous
section we obtain
\begin{equation}
T_{\lambda }=A\frac{\hbar ^{2}}{m}n^{2/3}\frac{\eta }{\eta
_{0}}F\left( \frac{1}{(1-x)^{2/3}}B\frac{\eta }{\eta _{0}}\right)
,\qquad \eta _{0}=\hbar n  \label{h3}
\end{equation}
where $\eta /\eta _{0}$ is again given by Eq. (\ref{pol}). The
universality hypothesis means that $\xi $ is independent of $x$,
so that we may consider $A$ and $B$ to be defined by Eqs.
(\ref{ab}) and (\ref{fit}). Then the only dependence of the
superfluid transition temperature on $x$ is through the explicit
factor of $(1-x)^{-2/3}$ in Eq. (\ref{h3}).

\begin{figure}[htp]
\epsfxsize=3.375in \centerline{\epsffile{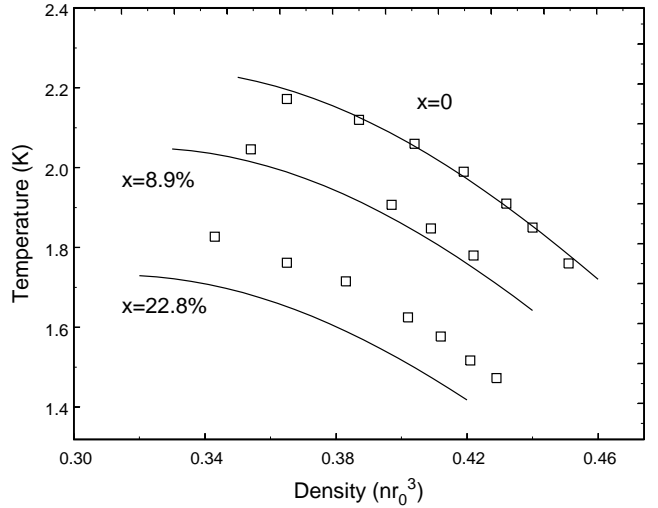}}

\caption{Critical temperature of $^{3}$He-$^{4}$He mixtures versus
the reduced density $n^{\ast}=nr_{0}^{3}$ ($r_{0}=2.556$ \AA) for
$^{3}$He concentrations $x=0$, 8.9\% and 22.8\%. Solid line is the
theory (\protect\ref{h3}) with $A$ and $B$ the same as for pure
$^{4}$He, while squares represent experimental data. Only the
$x=0$ curve was fitted to experiment.}

\label{fig4}
\end{figure}

To check the validity of the assumptions made we are to compare
the formula (\ref{h3}) with experimental data. Here we make use of
experiments \cite{LePair} where $T_{\lambda }$ was measured in
$^{3}$He-$^{4}$He mixtures under pressure at several values of
$x$. We transform these results into $T_{\lambda }(n)$ dependence
using equations of state for different $x$ at $T=1.5$ K, which
hardly can introduce any significant error. These data are shown
by squares in Fig. 4 for $x=0.089$ and $x=0.228$ along with the
pure $^{4}$He results. Solid curves represent $T_{\lambda }$ as
obtained from Eq. (\ref {h3}). Note that only the $x=0$ curve was
fitted to experiment.

We see that qualitatively the formula (\ref{h3}) correctly
describes the dependence of the transition temperature both on
density and on $x$, though for larger $x$'s the theoretical curves
tend to lie lower than the experimental points. It is difficult to
decide unambiguously whether this deviation indicate possible weak
dependence of  $\xi $ on $x$ or simply the approximations made in
evaluating $R^2$ are too crude. In any case the estimate
(\ref{h3}) makes sense only for small $x$.

\section{Conclusions}

In this paper the Lindemann-like superfluidity criterion proposed
earlier \cite{ap} was applied to several systems where the
superfluid transition is known to occur, namely to a weakly
non-ideal Bose gas,  liquid helium and  $^{3}$He-$^{4} $He
mixtures. The mean square displacement $R^2$ of a particle in
imaginary time which enters in the criterion, and hence
$T_{\lambda }$ were related here to the zero frequency friction
coefficient (or to the shear viscosity in case of liquids). We
found that in all cases the results obtained do qualitatively
reproduce the observed behaviour of the transition temperature
$T_{\lambda }$, as obtained from experiment (including numerical
modeling).

As far as the quantitative accuracy of the approach is concerned
the situation is still not fully understood. It is clear, however,
that one can hardly expect it to be very accurate. Indeed, the
criterion proposed may be of real interest only if the critical
temperature differs significantly from the ideal gas value (in the
opposite case the estimate (\ref{id}) is quite sufficient). But it
is obvious e.g. from Eq. (\ref{exp}), valid at $\hbar \gamma
/T_{0}\gg 1$, that the result in this case is rather sensitive to
the exact values of the damping parameter $\gamma $ and the
phenomenological parameter $\xi$ since they both stand in the
exponential. These parameters are in general not known exactly and
even a small variation in $\gamma $ or $\xi$ may significantly
influence the resulting estimate for $T_{\lambda }$.

It seems therefore that the best use of the formulas obtained may
be the analysis of how the transition temperature, which e.g. is
known for a given system, may change if some parameters of the
system are slightly modified. In this case one can fix all
adjustable parameters by fitting to the known data and then
analyze their deviations for a modified system. A simple example
of how this can be done was given in Sec. V where
$^{3}$He-$^{4}$He mixtures were discussed. We have first fixed all
free parameters comparing the theory to the pure $^{4}$He data and
then take account of the nonzero concentration of $^{3}$He atoms.

The criterion proposed leads also to an unexpected prediction for
the weakly interacting Bose gas. In this case the initial increase
of $T_{\lambda }$ with interaction strength may be attributed to
the change of the Lindemann parameter $\xi $ and this results in
an upper bound on the peak value of $T_{\lambda }/T_{\lambda
}^{0}$. Indeed, in the limit of zero density $\xi \simeq 0.15$ and
at helium densities we have an estimate $\xi \sim 0.12\div 0.13$
(see Ref. \onlinecite{ap} and Sec. IV) so that the change in $\xi
$ and hence the maximum value of $T_{\lambda }/T_{\lambda }^{0}$
could not be large. In Sec. III we obtain a rough estimate
$T_{\lambda }/T_{\lambda }^{0}<1.2$ consistent with the PIMC
results \cite{Gruter}. Possible numerical evaluation of $\xi $ in
a dense system can make this inequality more accurate.

Numerical estimates of $\xi$ would be desirable also from the
general point of view. This is practically the only way to make a
final decision concerning the validity of the approach presented
here. One has to calculate $R^2$ along the $\lambda$ line and then
check whether $\xi=R^2/a^2$ can be regarded as a constant (at
least in some density range) and to what extent is $\xi$ universal
in systems with e.g. different interaction potentials.

I am grateful to P.I. Arseyev and V.V. Losyakov for usefull
discussions and to D.M. Ceperley for stimulating correspondence
concerning the Bose gas. This work was supported in part by
the Russian Foundation for the Fundamental Research (Project No
00-15-96698).

\end{document}